\documentclass[11pt, epsfig]{article}
  \usepackage{amsthm,amsfonts}
  \usepackage{amsmath}
\newcommand{\bea}   {\begin{eqnarray}}
\newcommand{\eea}   {\end{eqnarray}}
\begin{document}
\renewcommand{\thefootnote}{\fnsymbol{footnote}}

\thispagestyle{empty}

\title{Critical scaling dimension of $D$-module representations of ${\cal N}=4,7,8$ Superconformal Algebras and constraints on Superconformal Mechanics}

\author{Sadi Khodaee\thanks{{\em e-mail: khodaee@cbpf.br}}
~and Francesco
Toppan\thanks{{\em e-mail: toppan@cbpf.br}}
\\
\\
}
\maketitle
\centerline{{\it CBPF, Rua Dr. Xavier Sigaud 150, Urca,}}{\centerline {\it\quad
cep 22290-180, Rio de Janeiro (RJ), Brazil.}}

\maketitle
\begin{abstract}
At critical values of the scaling dimension $\lambda$, supermultiplets of the global ${\cal N}$-Extended one-dimensional Supersymmetry algebra induce
$D$-module representations of finite superconformal algebras (the latters being identified in terms of the global supermultiplet and its critical scaling dimension). \par 
For ${\cal N}=4,8$ and global supermultiplets $(k, {\cal N}, {\cal N}-k)$, the exceptional superalgebras $D(2,1;\alpha)$ are recovered for ${\cal N}=4$, with a relation between $\alpha$ and the scaling dimension given by $\alpha= (2-k)\lambda$.
For ${\cal N}=8$ and $k\neq 4$ all four ${\cal N}=8$ finite superconformal algebras are recovered, at the critical values $\lambda_k = \frac{1}{k-4}$, with the following identifications:
$D(4,1)$ for $k=0,8$,  $F(4)$ for $k=1,7$, $A(3,1)$ for $k=2,6$ and $D(2,2)$ for $k=3,5$.
\par
The ${\cal N}=7$ global supermultiplet $(1,7,7,1)$ induces, at $\lambda= -\frac{1}{4}$, a $D$-module representation of the exceptional superalgebra $G(3)$.\par 
$D$-module representations are applicable to the construction of superconformal mechanics in a Lagrangian setting. The isomorphism of the $D(2,1;\alpha)$ algebras under an $S_3$ group action on $\alpha$, coupled with the relation between $\alpha$ and the scaling dimension $\lambda$, induces non-trivial constraints on the admissible models of ${\cal N}=4$ superconformal mechanics. The existence of new superconformal models is pointed out. E.g., coupled $(1,4,3)$ and $(3,4,1)$ supermultiplets generate an ${\cal N}=4$ superconformal mechanics if $\lambda$ is related to the golden ratio. \par
The relation between classical versus quantum $D$-module representations is presented.
\end{abstract}
\vfill

\rightline{CBPF-NF-012/12}

\newpage
\section{Introduction}
In this work we extend the construction of \cite{kt0} proving that all finite ${\cal N}=7,8$ superconformal algebras are recovered as $D$-module representations for a critical value
of the scaling dimension $\lambda$ of some given global supermultiplet. Therefore, together with the results concerning ${\cal N}=4$, a criticality is encountered for $D$-module representations of the ${\cal N}=4,7,8$ finite superconformal algebras. This criticality has deep consequences in constructing and constraining the admissible
one-dimensional superconformal mechanical models in a Lagrangian setting.\par
In one dimension the superconformal invariance can be characterized by the superconformal
algebras, which are recovered from the list of finite simple Lie superalgebras \cite{{kac},{nrs},{dictionary},{VP}} with further restrictions. Their even sector ${\cal G}_{even}$ is a direct sum
${\cal G}_{even}= sl(2)\oplus R$ (where $R$ is known as the $R$-symmetry), while the odd sector
${\cal G}_{odd}$ is spanned by $2{\cal N}$ odd generators (${\cal N}$ denotes the value of the extended supersymmetry). The superalgebras are closed under (anti)commutators and satisfy the
graded Jacobi identities.  The $sl(2)$ generators will be denoted as $D,K,H$ ($D$ is the dilatation
operator). Explicitely, they satisfy the commutation relations
\bea\label{sl2alg}
&\relax [D,H] =H,\quad\quad [D,K]=-K,\quad\quad [H,K] =2 D.
\eea
A grading is induced by the dilatation operator $D$, so that (${\cal G}_i$ is the sector of grading $i$ and for any $g_i^\alpha\in {\cal G}_i$ the commutator $[D,g_i^\alpha]=ig_i^\alpha$ holds)
\bea
{\cal G}_{even} = {\cal G}_{-1}\oplus 
{\cal G}_0\oplus {\cal G}_{1}, \quad &&\quad
{\cal G}_{odd} =  {\cal G}_{-\frac{1}{2}}\oplus {\cal G}_{\frac{1}{2}}.
\eea 
The sector ${\cal G}_1$ (${\cal G}_{-1}$) contains a unique generator given by $H$ ($K$). The odd sectors ${\cal G}_{\frac{1}{2}}$ and ${\cal G}_{-\frac{1}{2}}$ are spanned by the ${\cal N}$ supercharges
$Q_I$'s and their ${\cal N}$ superconformal partners ${\widetilde Q}_I$'s, respectively. The ${\cal G}_0$ sector is given by the union of $D$ and the $R$-symmetry subalgebra (${\cal G}_0={\{D\}}\bigcup\{R\}$). \par
We have, in particular, that the following anticommutation relations are satisfied for $I,J=1,2,\ldots, {\cal N}$:
\bea
\{Q_I,Q_J\} &=& 2\delta_{IJ} H,\nonumber\\
\{{\widetilde Q}_I,{\widetilde Q}_J\}&=& - 2 \delta_{IJ}K,
\eea
with furthermore
\bea
\relax &[H,Q_I]=[K,{\widetilde Q}_I]=0.&
\eea
It follows that the positive sector ${\cal G}_{>0}={\cal G}_{\frac{1}{2}}\oplus {\cal G}_1$, spanned by the generators $Q_I$'s and $H$, is isomorphic to the one-dimensional, ${\cal N}$-extended,
global supersymmetry, namely the superalgebra underlying the Supersymmetric Quantum Mechanics.\par
The $sl(2)$ algebra admits a (non-critical) $D$-module representation, expressed by the differential operators 
\bea\label{sl2dmod}
H&=&\frac{d}{dt},\nonumber\\
D&=&-t\frac{d}{dt}-\lambda,\nonumber\\ 
K&=&-t^2\frac{d}{dt} -2\lambda t.
\eea
This representation is non-critical since it closes the (\ref{sl2alg}) $sl(2)$ algebra for any value
of the scaling dimension $\lambda$.\par
On the other hand, linear $D$-modules representations for the ${\cal N}$-extended global supersymmetry algebra \cite{witten} spanned by the generators $Q_I$'s and $H$ have been intensively studied
in the last decade \cite{pt}-\cite{gikt}. They are surprisingly rich and intricated. They can be divided into several classes: minimal versus non-minimal representations, homogeneous versus inhomogeneous representations, admitting (or not, see \cite{gikt}) a graphical presentation, etc.
Those features which are relevant for the present work will be briefly recalled in Appendix {\bf A}.\par 
Following \cite{kt0}, we require the compatibility condition of the (\ref{sl2dmod}) $sl(2)$ $D$-module representations with the $D$-module representations of the ${\cal N}$-extended global
supersymmetries. The $H,D,K$ generators are assumed to act diagonally on the component fields
of the global supermultiplets. The relative scaling dimension of the component fields entering the global supermultiplets are unambiguously fixed by the dimensionality ($[Q_I]=\frac{1}{2}, [H]=1$) of the generators of the global supersymmetry. A unique free parameter is left. It is the overall scaling dimension $\lambda$ of the global supermultiplet which, by definition, coincides with the lowest scaling dimension of its component fields.\par
The finite superconformal algebras introduced above admit the following properties. Their generators can be recovered by repeatedly applying the (anti)commutation relations involving 
the supercharges $Q_I$ and the conformal generator $K$. 
We have therefore the possibility to check, for any given global supermultiplet  with $2n$ component fields and any overall
assignment $\lambda$ of the scaling dimension, whether the Ansatz for $K$
expressed by 
\bea\label{Kgenerator}
K&=&-t^2\frac{d}{dt}{\bf 1}_{2n}-2t\Lambda
\eea
($\Lambda$ is a diagonal matrix
whose diagonal entries coincide with the scaling dimensions of the component fields so that, for a length-$l$ supermultiplet, see Appendix {\bf A}, $\Lambda$ is given by $\Lambda = diag(\lambda_1=\lambda,\ldots, \lambda_{2n} =\lambda+\frac{l-1}{2})$)  induces
a $D$-module representation for a finite superconformal algebra (to be determined).  \par
To this end we recall that in a finite superconformal algebra the conformal superpartners ${\widetilde Q}_I$ of the global supercharges $Q_I$ are recovered through the position
\bea\label{Qtildes}
\relax [K, Q_I] &=& {\widetilde Q}_I,
\eea
which can be assumed to be the definition of the ${\widetilde Q}_I$ generators. The closure
of the superalgebra requires the introduction of new even generators
\bea\label{Sijgenerators}
S_{IJ} &=& \{ Q_I, {\widetilde Q}_J\}.
\eea
By making use of the Jacobi identities one can easily prove that, for any $I$, $S_{II}=-2D$ while, for $I\neq J$, the antisymmetric property $S_{IJ}=-S_{JI}$ holds. For $I>J$ the $S_{IJ}$'s generators commute with $H$ and $K$,
\bea
[H,S_{IJ}]=[K,S_{IJ}]=0,   \quad (I>J),
\eea
so that they can be identified with the $R$-symmetry generators. The $S_{IJ}$ generators are not, in general, linearly independent.\par
The explicit construction of the differential operators ${\widetilde Q}_I$ and $S_{IJ}$ from repetead (anti)-commutators involving $K$ and the $Q_I$'s does not yet guarantee that we have 
a $D$-module representation for a finite ${\cal N}$-extended superconformal algebra. We need
to verify that the superalgebra of differential operators $H,K,D, Q_I, {\widetilde Q}_I, S_{IJ}$ closes
without the introduction of further generators. For this purpose it is sufficient to check whether
the commutators involving the $S_{IJ}$'s generators and the $Q_K$'s close on the global supercharges, namely that
\bea\label{closurecond}
\relax [S_{IJ}, Q_K] &=& \alpha_{IJ,K}^LQ_L
\eea
is verified for some structure constants $\alpha_{IJ,K}^L$. Indeed,
by making use of the Jacobi identities the (\ref{closurecond}) equation implies the closure
of the commutators $[S_{IJ}, {\widetilde Q}_K]$ and $[S_{IJ}, S_{MN}]$. We have
\bea
\relax[S_{IJ}, {\widetilde Q}_K] &=& \alpha_{IJ,K}^L{\widetilde Q}_L ,\nonumber\\
\relax [S_{IJ}, S_{MN}] &=& -\{Q_M,[{\widetilde Q}_N, S_{IJ}]\}+\{{\widetilde Q}_N, [S_{IJ}, Q_M]\}=
 \alpha_{IJ,N}^L S_{ML}+ \alpha_{IJ,M}^LS_{LN}.
\eea
The condition (\ref{closurecond}) on the differential operators will be called the ``closure condition".  In the cases that we investigated three possibilities are encountered, associated with the choice of the global supermultiplet and its scaling dimension $\lambda$:\\
{\em i}) the closure condition is automatically satisfied for any value of $\lambda$ (no criticality in this case),\\
{\em ii}) the closure condition leads to a linear equation for $\lambda$ which pinpoints the
critical value $\lambda$,\\
{\em iii}) the closure condition admits no solution. In this latter case the global supermultiplet cannot be lifted to a $D$-module representation for a finite superconformal algebra.\par
To perform the needed computations we developed a package for Mathematica. We report here the results. 
All four ${\cal N}=8$ finite superconformal algebras are recovered from a $D$-module representation induced by the ${\cal N}=8$ $(k,8,8-k)$ global supermultiplets (see Appendix {\bf A}) for $k\neq 4$ at the critical values $\lambda_k = \frac{1}{k-4}$, with the following identifications:
$D(4,1)$ for $k=0,8$,  $F(4)$ for $k=1,7$, $A(3,1)$ for $k=2,6$ and $D(2,2)$ for $k=3,5$.
For $k=4$ the closure condition admits no solution. Furthermore, the unique 
${\cal N}=7$ global supermultiplet which cannot be extended to an ${\cal N}=8$ supermultiplet (identified by its field content $(1,7,7,1)$ and first introduced in \cite{krt}) induces, at the critical value $\lambda= -\frac{1}{4}$, a $D$-module representation of the exceptional superalgebra $G(3)$.\par 
For ${\cal N}=4$ the class of exceptional superalgebras $D(2,1;\alpha)$, parametrized by $\alpha$, is recovered for the $(k,4,4-k)$ global supermultiplets with the identification 
$\alpha= (2-k)\lambda$ in terms of the scaling dimension $\lambda$. The unique global ${\cal N}=3$ supermultiplet which cannot be extended to an ${\cal N}=4$ supermultiplet (identified by its field content $(1,3,3,1)$) induces the $D$-module of the $B(1,1)=osp(3|2)$ ${\cal N}=3$
superconformal algebra for any value of the scaling dimension $\lambda$ (no criticality).
\par
Besides these results we also proved that the inhomogeneous extension of the ${\cal N}=8$ global supermultiplet $(3,8,5)$ induces a $D$-module representation for the $D(2,2)$ superalgebra. The inhomogeneous term is essential, see \cite{{lt},{kt0}}, to introduce Calogero-type terms \cite{dff} in superconformal mechanics \cite{fr}-\cite{fil}. We further analyzed a few cases of ${\cal N}=6$ global supermultiplets and their induced $D$-module representations.\par
We reviewed the construction of superconformal mechanics (in a Lagrangian setting) recovered from the $D$-module representations of the finite superconformal algebras. As an example, we
proved that the ${\cal N}=8$ global action associated to the $(1,8,7)$ supermultiplet, under a homogeneity condition and in presence of a non-trivial interaction (see (\ref{187model}) and the following discussion), is invariant under the exceptional $F(4)$ superalgebra. We thus recovered, in a different framework, the finding of \cite{di}. The previously cited ${\cal N}=8$ inhomogeneous $(3,8,5)$ supermultiplet induces, on the other hand,  a new $D(2,2)$-invariant superconformal mechanical model that will be presented elsewhere. \par
In application to classical superconformal mechanics in a Lagrangian framework the scaling dimensions of the component fields have to satisfy a reality condition. On the other hand the
$D$-module operators can be applied to quantum systems. In this case they should satisfy a
Hermiticity condition which depends on a chosen metric $\eta$. For the dilatation operator $D$ the chosen metric can be constant (either $1$ or depending on the momentum operator $p_t$)
or non-constant. It is well-known that, by assuming the metric to be $1$, the Hermiticity condition of the dilatation operator implies that the scaling dimension $\lambda$ should belong 
to the critical strip $\lambda=\frac{1}{2}+i\gamma$, with $\gamma\in \Bbb{R}$. This is the critical strip where the non-trivial zeros of the Riemann's zeta function are encountered
(the Hermiticity property of $D$ is at the basis of an attempt \cite{BK} to prove the Riemann hypothesis). We linked the Hermiticity constraints on the scaling dimensions with the choice  (constant and non-constant) of the metric $\eta$.
\par
The last part of the paper is devoted to a thourough investigation of the constraints on ${\cal N}=4$ superconformal mechanics resulting from the ${\cal N}=4$ criticality condition. Multi-particle superconformal mechanics is based on several interacting supermultiplets which carry a representation of the same finite superconformal algebra. The ${\cal N}=4$ exceptional
superconformal algebras $D(2,1;\alpha)$ are isomorphic for values of $\alpha$ which are related
by an $S_3$ group of transformations (see (\ref{alphas})). This fact, together with the critical relations between $\alpha$ and the scaling dimension $\lambda$ for the various global ${\cal N}=4$ supermultiplets, has deep and non-trivial consequences in constraining multiparticle superconformal mechanics. The origin of these constraints are of representation theoretical
nature. We derived in particular, see Appendix {\bf B}, the admissible common scaling dimensions $\lambda$ which allow inequivalent global  ${\cal N}=4$ supermultiplets to induce $D$-module representations for the same superconformal algebra. As an application we find that in certain cases irrational solutions for $\lambda$ exist. The superconformal models based on these
interacting supermultiplets are ${\cal N}=4$ invariant, but cannot be extended to a full ${\cal N}=8$ invariance. One particular superconformal example, obtained from the (\ref{goldenprep}) prepotential,  involves the interaction of the $(1,4,3)$ and the $(3,4,1)$
supermultiplets and is based on a scaling dimension related to the golden mean.\par
The representation theoretical nature of the ${\cal N}=4$ constraints has implications for the
critical scaling dimensions of the ${\cal N}=7,8$ superconformal algebras $D$-modules. 
This is due to the fact that the minimal ${\cal N}=7,8$ supermultiplets admit (at least) one
decomposition in terms of two minimal ${\cal N}=4$ supermultiplets. The critical scaling dimensions can be sometimes partly and sometimes completely determined (as it indeed happens
for the ${\cal N}=8$ $(5,8,3)$ and $(3,8,5)$ $D$-modules) by the ${\cal N}=4$ analysis. In Appendix {\bf B} it is shown how the ${\cal N}=4$ constraints imply the absence of ${\cal N}=8$ superconformal algebras induced by the $(4,8,4)$ supermultiplet.
\par
The scheme of the paper is the following. In Section {\bf 2} we review the results of \cite{kt0}
about $D$-module representations of the ${\cal N}=4$ superconformal algebras from minimal
global ${\cal N}=4$ supermultiplets and we present the derivation of the non-critical $D$-module
representation of the $B(1,1)$ simple superalgebra from the ${\cal N}=3$ $(1,3,3,1)$ global supermultiplet. In Section {\bf 3} the four finite ${\cal N}=8$ superconformal algebras are realized as $D$-module representations from the ${\cal N}=8$ global supermultiplets 
$(k,8,8-k)$ for critical values of the scaling dimension $\lambda$ associated with $k\neq 4$. The unique ${\cal N}=7$
finite superconformal algebra $G(3)$ is recovered, at the critical value $\lambda=-\frac{1}{4}$, from the $(1,7,7,1)$ supermultiplet, namely the unique ${\cal N}=7$ minimal supermultiplet which cannot be extended to an ${\cal N}=8$ representation. Other cases, both critical and non-critical, of $D$-module representations for ${\cal N}=6$ superconformal algebras are presented in Section {\bf 4}. In Section {\bf 5} we analyze the Hermiticity conditions of the $D$-module representations in association with a constant or non-constant metric. In Section {\bf 6} we
review the construction of the superconformal mechanics in a Lagrangian setting derived from $D$-module representations of superconformal algebras. In Section {\bf 7} the constraints on (multiparticle) superconformal mechanics are derived from the criticality condition of ${\cal N}=4$ superconformal algebras and the isomorphism of the $D(2,1;\alpha)$ superalgebras under the $S_3$ group of transformations acting on the parameter $\alpha$.  The existence of
${\cal N}=4$ multi-particle superconformal mechanics for certain irrational values of the scaling dimension $\lambda$ of the supermultiplets is pointed out (in the Appendix the admissible real values $\lambda$, associated to pairs of ${\cal N}=4$ supermultiplets which carry a $D$-module representation for the same ${\cal N}=4$ superconformal algebra, are explicitly presented). In the Conclusions we comment about these results and discuss possible future
applications.

\section{$D$-module representations of ${\cal N}=3,4$ Superconformal algebras}

We summarize at first the results of \cite{kt0} concerning the $D$-module representations of finite ${\cal N}=4$ superconformal algebras induced by the minimal linear representations
of the ${\cal N}=4$ global supersymmetry. We postpone to Section {\bf 7} a detailed discussion of the finite ${\cal N}=4$ superconformal algebras and the interpretation of the results here reported. \par
The minimal homogeneous linear global ${\cal N}=4$ supermultiplets are expressed in terms of their field content  $(k,4,4-k)$, with $k=0,1,2,3,4$. The $k=0,1$ global supermultiplets admit a inhomogeneous extension \cite{kt0}.   \par
For $k=2$, the $D$-module representation of the $A(1,1)=sl(2|2)/{\cal Z}$ superalgebra is encountered at the
$\lambda=0$ value of the scaling dimension, while a $D$-module representation of $sl(2|2)$ is found at $\lambda\neq 0$.\par
For $k\neq 2$, $D$-module representations of the ${\cal N}=4$ exceptional superalgebras
$D(2,1;\alpha)$ are found. The identification between $\alpha$ and the scaling dimension 
is expressed by
\bea\label{alphalambda}
\alpha &=& (2-k)\lambda,
\eea
where the $A(1,1)$ superalgebra is recovered as a degenerate case for $\alpha=0,-1$
(see Section {\bf 7}).\par
The $(0,4,4)_{inhom}$ inhomogeneous extension of the $k=0$ global supermultiplet does not
induce $D$-module representations for finite ${\cal N}=4$ superconformal algebras. The $(1,4,3)_{inhom}$
inhomogeneous extension of $k=1$, on the other hand, induces a $D$-module representation
of the $A(1,1)$ superalgebra. The presence of the inhomogeneous term in the supertransformations forces the scaling dimension of the $(1,4,3)_{inhom}$ inhomogeneous supermultiplet to be given  by $\lambda=-1$. Since formula (\ref{alphalambda}) continues to hold, the $\alpha=-1$ value
is recovered.\par
At this stage we have already encountered, for ${\cal N}=4$, the phenomenon of ``critical scaling". Indeed, due to (\ref{alphalambda}), inequivalent $D(2,1;\alpha)$ superalgebras are identified for different
values of $\lambda$ (and $k$). It is worth mentioning that, for $\alpha\in {\Bbb C}\backslash\{0,-1\}$,
two $D(2,1;\alpha)$ superalgebras are isomorphic if and only if their parameters are related by
one of the six transformations belonging to the $S_3$ group given by (\ref{alphas}).\par
In Section {\bf 7} we will discuss the implications of the ${\cal N}=4$ critical scaling for
$D$-module representations of the ${\cal N}>4$ superconformal algebras and in constraining 
the Lagrangian formulation of ${\cal N}=4$ superconformal theories.\par
There exists a unique global ${\cal N}=3$ supermultiplet which cannot be extended to an ${\cal N}=4$ supermultiplet (see \cite{pt}). Its field content is $(1,3,3,1)$ and its component fields
will be denoted as $x, \psi_i, g_i,\omega$ ($i=1,2,3$). Their scaling dimensions are, respectively,
$[x]=\lambda$, $[\psi_i]=\lambda+\frac{1}{2}$, $[g_i]=\lambda+1$, $[\omega]=\lambda+\frac{3}{2}$ ($\lambda$ is the scaling dimension of the supermultiplet). The fields $x, g_i$ are assumed to be bosonic, while $\psi_i, \omega$ are assumed to be fermionic. They can be rearranged in the supermultiplet  $|m>$ such that
$|m>^T=<m|= <x, g_1,g_2,g_3;\omega,\psi_1,\psi_2,\psi_3|$. The three global supercharges
acting on $|m>$ are given by the $8\times 8$ supermatrices $Q_i$ obtained from
the ``root" supercharges ${Q_i^R}$ by applying the dressing transformation $S=
diag(1,\partial_t,\partial_t,\partial_t,\partial_t,1,1,1)$. \par
The $(1,3,3,1)$ supermultiplet induces a $D$-module representation of the ${\cal N}=3$ superconformal algebra $B(1,1)=osp(3,2)$, with $6$ even and $6$ odd generators. Its bosonic subalgebra is $sl(2)\oplus so(3)$. The $D$-module representation is obtained for any $\lambda$,
since the ``closure condition" (\ref{closurecond}) gives no restriction on $\lambda$ (there is no critical scaling 
in this case). The construction outlined in the Introduction produces, besides the generators $H, D,K$ and $Q_i$'s, the superconformal partners ${\widetilde Q_i}$'s and the three independent $R$-symmetry generators $S_{ij}$, for $i>j$.\par
An explicit presentation of the $D$-module generators of $B(1,1)$ is given by the following supermatrices acting on $(4|4)$ supermultiplets (here and in the following $E_{mn}$ denotes
the matrix with entry $1$ at the crossing of the $m$-th row and $n$-th column and $0$
otherwise)
\bea\label{1331}
H&=& {\bf 1}_8\cdot\partial_t,\nonumber\\
D&=& -{\bf 1}_8\cdot t\partial_t -\Lambda,\nonumber\\
K&=& -{\bf 1}_8\cdot t^2\partial_t -2t\Lambda,\nonumber\\
Q_1&=&  (-E_{38}+E_{47}-E_{52}+E_{61})\partial_t+
E_{16}-E_{25}+E_{74}-E_{83}
,\nonumber\\
Q_2&=& (E_{28}-E_{46}-E_{53}+E_{71})\partial_t+E_{17}-E_{35}-E_{64}+E_{82},\nonumber\\
Q_3&=& (-E_{27}+E_{36}-E_{54}+E_{81})\partial_t+E_{18}-E_{45}+E_{63}-E_{72},\nonumber\\
{\widetilde Q}_1 &=&(-E_{38} +E_{47}-E_{52}+E_{61})t\partial_t+
(E_{16}-E_{25}+E_{74}-E_{83})t+\nonumber\\&&-
E_{52}(2\lambda+2)+(-E_{38}+E_{47})(2\lambda+1)+E_{61}2\lambda,
\nonumber\\
{\widetilde Q}_2 &=&(E_{28} -E_{46}-E_{53}+E_{71})t\partial_t+
(E_{17}-E_{35}-E_{64}+E_{82})t+\nonumber\\&&-
E_{53}(2\lambda+2)+(E_{28}-E_{46})(2\lambda+1)+E_{71}2\lambda,
\nonumber\\
{\widetilde Q}_3 &=&(-E_{27} +E_{36}-E_{54}+E_{81})t\partial_t+
(E_{18}-E_{45}+E_{63}-E_{72})t+\nonumber\\&&-
E_{54}(2\lambda+2)+(-E_{27}+E_{36})(2\lambda+1)+E_{81}2\lambda,
\nonumber\\
S_1&=&E_{34}-E_{43}+E_{78}-E_{87},\nonumber\\
S_2&=&-E_{24}+E_{42}-E_{68}+E_{86},\nonumber\\
S_3&=&E_{23}-E_{32}+E_{67}-E_{76},\nonumber\\
\eea
where $\Lambda$ is the diagonal matrix $\Lambda= diag(\lambda,\lambda+1,\lambda+1,\lambda+1,\lambda+\frac{3}{2},\lambda+\frac{1}{2},\lambda+\frac{1}{2},\lambda+\frac{1}{2})$
and $S_i =\epsilon_{ijk}S_{jk}$ ($\epsilon_{123}=1$).
\par
In this basis the (anti)commutation relations of the $B(1,1)$ superalgebra reads as
\bea\label{b11}
&
\begin{array}{llllll}
[H,K]&=&2D,~&&&\\
\relax [D,H]&=&H,~&[D,K]&=&-K, \\
\relax [D,Q_i]&=&\frac{1}{2} Q_i,~&[D,{\widetilde Q}_i]&=&-\frac{1}{2}{\widetilde Q}_i,\\
\relax [H,{\widetilde Q}_i]&=&Q_i,~& [K,{Q}_i]&=&{\widetilde Q}_i , \\
\{Q_i,Q_j\}&=&2\delta_{ij}H, ~& \{{\widetilde Q}_i, {\widetilde Q}_j\}&=&-2\delta_{ij}K, \\
\{Q_i,{\widetilde Q}_j\}&=&-2\delta_{ij} D+\epsilon_{ijk}S_k,~&[S_i,Q_j]&=&-\epsilon_{ijk}Q_k,\\
\relax [S_i,{\widetilde Q}_j]&=&-\epsilon_{ijk}{\widetilde Q}_k,~&[S_i,S_j]&=&-\epsilon_{ijk}S_k.
\end{array}
&
\eea
With this ${\cal N}=3$ derivation we have shown explicitly our construction. In the following, in
order to avoid reproducing too cumbersome formulas, we limit ourselves to report the main results.\par
It could be instructive to close this Section with an example of a derivation of an ${\cal N}=4$
superconformal algebra from a non-minimal global ${\cal N}=4$ supermultiplet (whose number of component fields is doubled with respect to the minimal supermultiplets, see \cite{gkt}).
The unique length-$5$ ${\cal N}=4$ supermultiplet is the ``enveloping supermultiplet" \cite{krt}, with field content $(1,4,6,4,1)$. It induces a $D$-module representation for any $\lambda$ (the (\ref{closurecond}) closure condition is automatically satisfied) and the derived
superalgebra is unique (it does not depend on the scaling dimension $\lambda$). It is given
by $D(2,1;\alpha=1)$. At this special value of $\alpha$, $D(2,1;1)$ is also denoted as
$D(2,1)$. It corresponds to a superalgebra which belongs to the $D(m,n)=osp(2m|2n)$ classical series \cite{dictionary}.  

\section{Critical scaling dimensions and $D$-module representations of ${\cal N}=7,8$ Superconformal Algebras}

The minimal homogeneous linear global ${\cal N}=8$ supermultiplets are expressed in terms of their field content  $(k,8,8-k)$, with $k=0,1,2,\ldots, 8$. The $k=0,1,2,3$ global supermultiplets admit a inhomogeneous extension \cite{lt}. There exists a unique minimal, global ${\cal N}=7$,
linear supermultiplet which cannot be extended to ${\cal N}=8$. It is a length-$4$ supermultiplet whose field content is $(1,7,7,1)$, see \cite{krt}. \par
Before reporting the results of their $D$-module induced representations for finite superconformal algebras we recall that,
over ${\Bbb C}$, there are four finite ${\cal N}=8$ superconformal algebras and one finite ${\cal N}=7$ superconformal algebra \cite{dictionary}. The finite ${\cal N}=8$ superconformal algebras are:
\\
{\em i}) the $A(3,1)=sl(4|2)$ superalgebra, possessing  $19$ even generators and bosonic sector given by $sl(2)\oplus sl(4)\oplus u(1)$,  \\
{\em ii}) the $D(4,1)=osp(8,2)$ superalgebra, possessing $31$ even generators and bosonic sector given by $sl(2)\oplus so(8)$,\\
{\em iii}) the $D(2,2)=osp(4|4)$ superalgebra, possessing $16$ even generators and bosonic
sector given by $sl(2)\oplus so(3)\oplus sp(4)$,\\
{\em iv}) the $F(4)$ exceptional superalgebra, possessing $24$ even generators and bosonic sector given by $sl(2)\oplus so(7)$.\par
The finite ${\cal N}=7$ superconformal algebra is the exceptional superalgebra $G(3)$, possessing $17$ even generators and bosonic sector given by $sl(2)\oplus g_2$.\par
We start with the global supercharges $Q_I$'s obtained by dressing (see (\ref{qdressing})) the root supercharges $Q_I^R$ given in (\ref{n8scharges}). The hamiltonian $H$ is $H=\partial_t\cdot {\bf 1}$, while
the conformal generator $K$ is given by formula (\ref{Kgenerator}). The diagonal matrix $\Lambda$ entering (\ref{Kgenerator}) is
unambiguously determined in terms of the dressing transformation $S$ and the overall scaling dimension $\lambda$. The remaining generators of the superconformal algebras (${\widetilde Q}_I, D, S_{IJ}$) are obtained from equation (\ref{Qtildes}) (${\widetilde Q}_I$) and (\ref{Sijgenerators}) ($D$ and $S_{IJ}$). The closure condition (\ref{closurecond}) admits no solution for the global ${\cal N}=8$
$(4,8,4)$ supermultiplet. In the remaining cases it fixes the critical value $\lambda$ of the scaling dimension. \par
The results of the $D$-module representations of the ${\cal N}=8$ finite superconformal algebras present a $(k,8,8-k)\leftrightarrow (8-k,8,k)$ duality which is already encountered 
in the global supersymmetry (see \cite{krt}).\par The results are the following:
\subsection{Critical $D$-module representations of $D(4,1)$ from $(8,8,0)$ at $\lambda=\frac{1}{4}$ and
$(0,8,8)$ at $\lambda=-\frac{1}{4}$.}
The respective diagonal dressing matrices $S$ (and their associated diagonal matrices $\Lambda$ with the scaling dimensions of the component fields) are explicitly given by
\bea
k=8&:&  S=diag(1,1,1,1,1,1,1,1,1,1,1,1,1,1,1,1),\nonumber\\
&& \Lambda = diag(\frac{1}{4},\frac{1}{4},\frac{1}{4},\frac{1}{4},\frac{1}{4},\frac{1}{4},\frac{1}{4},\frac{1}{4},\frac{3}{4},\frac{3}{4},\frac{3}{4},\frac{3}{4},\frac{3}{4},\frac{3}{4},\frac{3}{4},\frac{3}{4}),\nonumber\\
k=0&:&  S=diag(\partial_t,\partial_t,\partial_t,\partial_t,\partial_t,\partial_t,\partial_t,\partial_t,1,1,1,1,1,1,1,1),\nonumber\\
&& \Lambda = diag(\frac{3}{4},\frac{3}{4},\frac{3}{4},\frac{3}{4},\frac{3}{4},\frac{3}{4},\frac{3}{4},\frac{3}{4},\frac{1}{4},\frac{1}{4},\frac{1}{4},\frac{1}{4},\frac{1}{4},\frac{1}{4},\frac{1}{4},\frac{1}{4}).
\eea
In both cases the $28$ $S_{IJ}$ generators for $I>J$ are all linearly independent (unambiguously identifying the finite superconformal algebra as $D(4,1)$). The non-vanishing values
of the structure constants $\alpha_{IJ,K}^L$ entering the closure condition (\ref{closurecond}) are
\bea
\alpha_{IJ,J}^I=1, && \alpha_{IJ,I}^J=-1.
\eea
For $k=8$ we recover the result of \cite{kt0}.
\subsection{Critical $D$-module representations of $F(4)$ from $(7,8,1)$ at $\lambda=\frac{1}{3}$ and
$(1,8,7)$ at $\lambda=-\frac{1}{3}$.}
The respective diagonal dressing matrices $S$ (and their associated diagonal matrices $\Lambda$ with the scaling dimensions of the component fields) are explicitly given by
\bea
k=7&:&  S=diag(1,1,1,1,1,1,1,\partial_t,1,1,1,1,1,1,1,1),\nonumber\\
&&\Lambda = diag(\frac{1}{3},\frac{1}{3},\frac{1}{3},\frac{1}{3},\frac{1}{3},\frac{1}{3},\frac{1}{3},\frac{4}{3},\frac{5}{6},\frac{5}{6},\frac{5}{6},\frac{5}{6},\frac{5}{6},\frac{5}{6},\frac{5}{6},\frac{5}{6}).\nonumber\\
k=1&:&  S=diag(1,\partial_t,\partial_t,\partial_t,\partial_t,\partial_t,\partial_t,\partial_t,1,1,1,1,1,1,1,1),\nonumber\\
&&\Lambda = diag(-\frac{1}{3},\frac{2}{3},\frac{2}{3},\frac{2}{3},\frac{2}{3},\frac{2}{3},\frac{2}{3},\frac{2}{3},\frac{1}{6},\frac{1}{6},\frac{1}{6},\frac{1}{6},\frac{1}{6},\frac{1}{6},\frac{1}{6},\frac{1}{6}).
\eea
In both cases $7$ relations reduce the number of linearly independent $S_{IJ}$ generators (for $I>J$) to $21$, unambiguously identifying the ${\cal N}=8$ finite superconformal algebra as $F(4)$.\par 
For $(7,8,1)$ we have
\bea
S_{21}-S_{65}-S_{74}+S_{83} &=& 0,\nonumber\\
S_{31}-S_{64}+S_{75}-S_{82} &=& 0,\nonumber\\
S_{32}+S_{54}+S_{76}+S_{81} &=& 0,\nonumber\\
S_{41}+S_{63}+S_{72}+S_{85} &=& 0,\nonumber\\
S_{42}-S_{53}-S_{71}+S_{86} &=& 0,\nonumber\\
S_{43}+S_{52}-S_{61}-S_{87} &=& 0,\nonumber\\
S_{51}+S_{62}-S_{73}-S_{84} &=& 0.
\eea
For $(1,8,7)$ we have
\bea
S_{21}-S_{65}+S_{74}-S_{83} &=& 0,\nonumber\\
S_{31}-S_{64}-S_{75}+S_{82} &=& 0,\nonumber\\
S_{32}+S_{54}-S_{76}-S_{81} &=& 0,\nonumber\\
S_{41}+S_{63}-S_{72}-S_{85} &=& 0,\nonumber\\
S_{42}-S_{53}+S_{71}-S_{86} &=& 0,\nonumber\\
S_{43}+S_{52}-S_{61}-S_{87} &=& 0,\nonumber\\
S_{51}+S_{62}+S_{73}+S_{84} &=& 0.
\eea
The structure constants of the superconformal algebra are explicitly computed. Their expression is too cumbersome to be explicitly reported here.

\subsection{Critical $D$-module representations of $A(3,1)$ from $(6,8,2)$ at $\lambda=\frac{1}{2}$ and
$(2,8,6)$ at $\lambda=-\frac{1}{2}$.}

The respective diagonal dressing matrices $S$ (and their associated diagonal matrices $\Lambda$ with the scaling dimensions of the component fields) are explicitly given by
\bea
k=6&:&  S=diag(1,1,1,1,1,1,\partial_t,\partial_t,1,1,1,1,1,1,1,1),\nonumber\\
&&\Lambda = diag(\frac{1}{2},\frac{1}{2},\frac{1}{2},\frac{1}{2},\frac{1}{2},\frac{1}{2},\frac{3}{2},\frac{3}{2},1,1,1,1,1,1,1,1).\nonumber\\
k=2&:&  S=diag(1,1,\partial_t,\partial_t,\partial_t,\partial_t,\partial_t,\partial_t,1,1,1,1,1,1,1,1),\nonumber\\
&&\Lambda = diag(-\frac{1}{2},-\frac{1}{2},\frac{1}{2},\frac{1}{2},\frac{1}{2},\frac{1}{2},\frac{1}{2},\frac{1}{2},0,0,0,0,0,0,0,0).
\eea
In both cases $12$ relations reduce the number of linearly independent $S_{IJ}$ generators (for $I>J$) to $16$, unambiguously identifying the ${\cal N}=8$ finite superconformal algebra as $A(3,1)$.\par 
For $(6,8,2)$ we have
\bea
S_{21}+S_{83} = 0, &&\quad S_{65}+S_{74} =0,\nonumber\\
S_{31}-S_{82} = 0, &&\quad S_{64}-S_{75} =0,\nonumber\\
S_{41}+S_{85} = 0, &&\quad S_{63}+S_{72} =0,\nonumber\\
S_{42}-S_{53} = 0, &&\quad S_{71}-S_{86} =0,\nonumber\\
S_{43}+S_{52} = 0, &&\quad S_{61}+S_{87} =0,\nonumber\\
S_{51}-S_{84} = 0, &&\quad S_{62}-S_{73} =0.
\eea
For $(2,8,6)$ we have
\bea
S_{21}-S_{83} = 0, &&\quad S_{65}-S_{74} =0,\nonumber\\
S_{31}+S_{82} = 0, &&\quad S_{64}+S_{75} =0,\nonumber\\
S_{41}-S_{85} = 0, &&\quad S_{63}-S_{72} =0,\nonumber\\
S_{42}-S_{53} = 0, &&\quad S_{71}-S_{86} =0,\nonumber\\
S_{43}+S_{52} = 0, &&\quad S_{61}+S_{87} =0,\nonumber\\
S_{51}+S_{84} = 0, &&\quad S_{62}+S_{73} =0.
\eea
The structure constants of the superconformal algebra are explicitly computed. Their expression is too cumbersome to be explicitly reported here.
\subsection{Critical $D$-module representations of $D(2,2)$ from $(5,8,3)$ at $\lambda=1$ and
$(3,8,5)$ at $\lambda=-1$.}
The respective diagonal dressing matrices $S$ (and their associated diagonal matrices $\Lambda$ with the scaling dimensions of the component fields) are explicitly given by
\bea
k=5&:&  S=diag(1,1,1,1,1,\partial_t,\partial_t,\partial_t,1,1,1,1,1,1,1,1),\nonumber\\
&&\Lambda = diag(1,1,1,2,2,2,2,2,\frac{3}{2},\frac{3}{2},\frac{3}{2},\frac{3}{2},\frac{3}{2},\frac{3}{2},\frac{3}{2},\frac{3}{2}).\nonumber\\
k=3&:&  S=diag(1,1,1,\partial_t,\partial_t,\partial_t,\partial_t,\partial_t,1,1,1,1,1,1,1,1),\nonumber\\
&&\Lambda = diag(-1,-1,-1,0,0,0,0,0,-\frac{1}{2},-\frac{1}{2},-\frac{1}{2},-\frac{1}{2},-\frac{1}{2},-\frac{1}{2},-\frac{1}{2},-\frac{1}{2}).
\eea
In both cases $15$ relations reduce the number of linearly independent $S_{IJ}$ generators (for $I>J$) to $13$, unambiguously identifying the ${\cal N}=8$ finite superconformal algebra as $D(2,2)$.\par 
For $(5,8,3)$ we have
\bea
S_{21}-S_{65}-S_{74}+S_{83} &=& 0,\nonumber\\
S_{31}+S_{64}-S_{75}-S_{82} &=& 0,\nonumber\\
S_{32}-S_{54}-S_{76}+S_{81} &=& 0,
\eea
together with
\bea
&&S_{41} = -S_{63} = S_{72} = -S_{85},\nonumber\\
&&S_{42} = S_{53} = -S_{71} = -S_{86},\nonumber\\
&&S_{43} = -S_{52} = S_{61} =-S_{87},\nonumber\\
&&S_{51} = S_{62} = S_{73} = S_{84}.
\eea
For $(3,8,5)$ we have
\bea
S_{21}-S_{65}+S_{74}-S_{83} &=& 0,\nonumber\\
S_{31}+S_{64}+S_{75}+S_{82} &=& 0,\nonumber\\
S_{32}-S_{54}+S_{76}-S_{81} &=& 0,
\eea
together with
\bea
&&S_{41} = -S_{63} = -S_{72} = S_{85},\nonumber\\
&&S_{42} = S_{53} = S_{71} = S_{86},\nonumber\\
&&S_{43} = -S_{52} = S_{61} =-S_{87},\nonumber \\
&&S_{51} = S_{62} = -S_{73} = -S_{84}.
\eea
The structure constants of the superconformal algebra are explicitly computed. Their expression is too cumbersome to be explicitly reported here.
\subsection{Critical $D$-module representations of $G(3)$ from $(1,7,7,1)$ at $\lambda=-\frac{1}{4}$.}
In this case the diagonal dressing matrix $S$ can be chosen, without loss of generality, to be
given by
\bea
S&=&diag( 1, \partial_t, \partial_t, \partial_t, \partial_t, \partial_t, \partial_t, \partial_t, \partial_t, 1, 1, 1, 1, 1, 1, 1).
\eea
After dressing, $7$ global supercharges $Q_I$'s ($I=1,\ldots, 7)$ remain as differential operators.
\par
$14$ out of the $21$ generators $S_{IJ}$ (for $I>J$) are linearly independent, due to the $7$
relations
\bea
S_{21}-S_{65}+S_{74} &=& 0,\nonumber\\
S_{31}-S_{64}-S_{75} &=& 0,\nonumber\\
S_{32}+S_{54}-S_{76} &=& 0,\nonumber\\
S_{41}+S_{63}-S_{72} &=& 0,\nonumber\\
S_{42}-S_{53}+S_{71} &=& 0,\nonumber\\
S_{43}+S_{52}-S_{61} &=& 0,\nonumber\\
S_{51}+S_{62}+S_{73} &=& 0.
\eea
At the critical scaling dimension $\lambda=-\frac{1}{4}$ one recovers the exceptional superalgebra $G(3)$.
\par
At this critical value the diagonal matrix $\Lambda$ with the scaling dimensions of the component fields is given by
\bea
\Lambda&=&diag( -\frac{1}{4},\frac{3}{4},\frac{3}{4},\frac{3}{4},\frac{3}{4},\frac{3}{4},\frac{3}{4},\frac{3}{4},\frac{5}{4},\frac{1}{4},\frac{1}{4},\frac{1}{4},\frac{1}{4},\frac{1}{4},\frac{1}{4},\frac{1}{4}).
\eea
\subsection{$D$-module representation of $D(2,2)$ from the inhomogeneous $(3,8,5)$ supermultiplet.}

There exists a unique inhomogeneous global supermultiplet which induces a $D$-module representation for a finite ${\cal N}=8$ superconformal algebra. The presence of the inhomogeneous extension fixes the scaling dimension of the auxiliary fields to be $0$ (see
\cite{kt0}). This implies that the overall scaling dimension of the supermultiplet has to be
$\lambda=-1$. The homogeneous $(3,8,5)$ supermultiplet 
is the only one inducing a $D$-module representation at this critical value of $\lambda$, the superconformal algebra being $D(2,2)$.
It remains to be checked whether the presence of the inhomogeneous extension is compatible
with the $D$-module representation for $D(2,2)$. The answer is positive. We adapted the construction discussed in \cite{kt0} for the $D$-module representation induced by the inhomogeneous $(1,4,3)$ supermultiplet. The eight supercharges act now on a $(9|8)$ supermultiplet
$|m>$ such that $<m|^T=(x_i, g_j, 1; \psi_a)$ ($i=1,2,3$,  $j=1,\ldots,5$, $a=1,\ldots, 8$).
They are explicitly given by
\bea
Q_1 &=& E_{1,11} -E_{2,10}-E_{3,13} +E_{12,4}+E_{14,6}-E_{15,5}-E_{16,8}+E_{17,7}+\nonumber\\&&
(E_{4,12}-E_{5,15}+E_{6,14}+E_{7,17}-E_{8,16}-E_{10,2}+E_{11,1}-E_{13,3})\partial_t,\nonumber\\
Q_2 &=& E_{1,12} +E_{2,13}-E_{3,10} -E_{11,4}+E_{14,7}+E_{15,8}-E_{16,5}-E_{17,6}+\nonumber\\&&
(-E_{4,11}-E_{5,16}-E_{6,17}+E_{7,14}+E_{8,15}-E_{10,3}+E_{12,1}+E_{13,2})\partial_t,\nonumber\\
Q_3 &=&
 E_{1,13} -E_{2,12}+E_{3,11} -E_{10,4}+E_{14,8}-E_{15,7}+E_{16,6}-E_{17,5}+
\nonumber\\&&
(-E_{4,10}-E_{5,17}+E_{6,16}-E_{7,15}+E_{8,14}+E_{11,3}-E_{12,2}+E_{13,1})\partial_t,\nonumber\\
Q_4 &=& E_{1,14} +E_{2,15}+E_{3,16} -E_{10,5}-E_{11,6}-E_{12,7}-E_{13,8}+E_{17,4}+cE_{17,9}\nonumber\\&&
(E_{4,17}-E_{5,10}-E_{6,11}-E_{7,12}-E_{8,13}+E_{14,1}+E_{15,2}+E_{16,3})\partial_t,\nonumber\\
Q_5 &=& E_{1,15} -E_{2,14}+E_{3,17} -E_{10,6}+E_{11,5}-E_{12,8}+E_{13,7}-E_{16,4}-c E_{16,9}+\nonumber\\&&
(-E_{4,16}+E_{5,11}-E_{6,10}+E_{7,13}-E_{8,12}-E_{14,2}+E_{15,1}+E_{17,3})\partial_t,\nonumber\\
Q_6 &=& E_{1,16} -E_{2,17}-E_{3,14} -E_{10,7}+E_{11,8}+E_{12,5}-E_{13,6}+E_{15,4}+c E_{15,9}+\nonumber\\&&
(E_{4,15}+E_{5,12}-E_{6,13}-E_{7,10}+E_{8,11}-E_{14,3}+E_{16,1}-E_{17,2})\partial_t,\nonumber\\
Q_7 &=& E_{1,17} +E_{2,16}-E_{3,15} -E_{10,8}-E_{11,7}+E_{12,6}+E_{13,5}-E_{14,4}-cE_{14,9}+\nonumber\\&&
(-E_{4,14}+E_{5,13}+E_{6,12}-E_{7,11}-E_{8,10}-E_{15,3}+E_{16,2}+E_{17,1})\partial_t,\nonumber\\
Q_8 &=& E_{1,10} +E_{2,11}+E_{3,12} +E_{13,4}+E_{14,5}+E_{15,6}+E_{16,7}+E_{17,8}+\nonumber\\&&
(E_{4,13}+E_{5,14}+E_{6,15}+E_{7,16}+E_{8,17}+E_{10,1}+E_{11,2}+E_{12,3})\partial_t.
\eea
One should notice the presence of the inhomogeneous constant $c$ in the $Q_4, Q_5,Q_6, Q_7$
transformations.
\par
The conformal generator $K$ (\ref{Kgenerator}) is uniquely specified by the diagonal matrix $\Lambda$ given by
\bea
\Lambda &=& diag(-1,-1,-1,0,0,0,0,0,0,-\frac{1}{2},-\frac{1}{2},-\frac{1}{2},-\frac{1}{2},-\frac{1}{2},-\frac{1}{2},-\frac{1}{2},-\frac{1}{2}).
\eea
The $D(2,2)$ superalgebra closes as in the homogeneous case (recovered for $c=0$).

\section{Some extra, critical and non-critical, $D$-modules with ${\cal N}=6$}

The strategy of searching for $D$-module representations of superconformal algebras requires, as an input, the supermultiplets of the global ${\cal N}$-Extended supersymmetry. For low values of
${\cal N}$ these supermultiplets have been classified. In these cases it is therefore possible to look for their possible, associated, superconformal $D$-modules. This paper is not the proper place for a systematic, exhaustive, investigation. We limit here to discuss some further selected cases,
besides the ones that we have already presented. To be definite, let us focus on the length-$4$ ${\cal N}=6$
global supermultiplets. Their resulting $D$-modules are of interest in the light of the analysis, based on their ${\cal N}=4$ decomposition,  discussed in Section {\bf 7} .\par
The three length-$4$ global ${\cal N}=6$ supermultiplets are \cite{krt} $(2,6,6,2)$, $(1,6,7,2)$ and
$(2,7,6,1)$. In all three cases we obtain a $D$-module representation for the $A(2,1)=sl(3|2)$
${\cal N}=6$ superconformal algebra, whose $R$-sector is the bosonic subalgebra $sl(3)\oplus u(1)$. The $15$ generators $S_{ij}$ (with $i,j=1,\ldots ,6$ and $i>j$) are not linearly independent. $6$ of them are determined in terms of the $9$ remaining generators, unambiguously identifying the $S_{ij}$'s as the $R$-sector of $A(2|1)$. The request that no
further odd generator is obtained, besides the $6$ global supercharges $Q_i$'s and their
$6$ conformal superpartners ${\widetilde Q}_i$'s, produces  no constraint on $\lambda$ for
the $(2,6,6,2)$ case (no critical scaling dimension). On the other hand the scaling dimensions
of $(1,6,7,2)$ and $(2,7,6,1)$ are constrained to the values, respectively, $\lambda=0$ and $\lambda=-\frac{1}{2}$. We can summarize these results in the following table. The $A(2,1)$  ${\cal N}=6$
superconformal algebra $D$-modules are recovered from
\bea
(2,6,6,2): &A(2,1),&\forall \lambda\in {\Bbb R},\nonumber\\
(1,6,7,2): &A(2,1),& for\quad \lambda=0,\nonumber\\
(2,7,6,1): &A(2,1),& for  \quad\lambda=-\frac{1}{2}.
\eea

\section{Classical versus quantum $D$-module representations}
 
The $D$-module representations of finite superconformal algebras that we introduced before
can be called ``classical representations".  Two equivalent viewpoints can be applied to their entries. They can be regarded either as differential operators in the variable $t$ (the ``time") or,
alternatively, they can be regarded as elements of an abstract Poisson brackets algebra generated 
by the relation $\{\pi_t, t\}=1$, where $\pi_t$ (which, as a differential operator, can be identified with
$\frac{d}{dt}$) is the conjugate momentum of $t$.\par
In Section {\bf 6} we present the construction of classical superconformal mechanics in a Lagrangian formalism from the classical $D$-module representations that we discussed so far.\par
The extension to quantum mechanics can be achieved in at least two different ways. The Lagrangian mechanics can be reformulated in the Hamiltonian framework, so that standard methods of quantization can be applied, at least in principle, to the classical Hamiltonian dynamics.\par
A more direct approach (the one we discuss here) consists in realizing the generators of the $D$-module representations as Hermitian operators. The entries will be expressed in terms of the Hermitian operators $t$ and
$p_t=i\frac{d}{dt}$.  We introduce at first the Hermitian generators for the $sl(2)$ diagonal subalgebra and the ${\cal N}$ global supercharges $Q_i$. The hermiticity properties of the remainining generators are determined as a consequence. It is convenient to express the Hermitian
$sl(2)$ generators $D,H,K$ acting on a given component field as
\bea\label{sl2her}
&H=p_t,\quad D= -(tp_t+i\lambda),\quad K=-(t^2p_t+2i\lambda t)&
\eea
 (the constraint on the scaling dimension $\lambda$ will be determined in the following), while
the ``quantum" $D$-module representation for the $Q_i$'s is obtained from the classical one by
replacing the $\pm \frac{d}{dt}$ entries ($\pm \frac{d}{dt}\rightarrow \pm p_t$), while leaving unchanged the $c$-number entries ($\pm 1$). We will see that this is the correct prescription to obtain Hermitian global supercharges.\par
For our purposes here it is sufficient to discuss the hermiticity properties of the dilatation operator $D$ and of the global supercharges $Q_i$. We require in particular that, acting on given supermultiplets $|m_j>$, the equalities 
\bea\label{hermit}
\int dt < m_1| \eta |Dm_2>= \int dt <Dm_1|\eta|m_2>, && 
 \int dt < m_1| \eta |Q_im_2>= \int dt <Q_im_1|\eta|m_2>,
\nonumber\\
\eea
(with $\eta$ a given metric to be specified) have to be satisfied. Let us discuss, for simplicity, the
$|m_1>=|m_2>\equiv |m>$ case and let us take $|m>$ as a $(k,{\cal N},{\cal N}-k)$ supermultiplet for ${\cal N}=4,8$ (the extension to other length-$3$ supermultiplets for arbitrary values of ${\cal N}$ is immediate). The component fields in the $|m>$ supermultiplets are
 $x_l$ ($l=1,\ldots, k$), $g_m$ ($m=1,\ldots , {\cal N}-k$) and the fermionic (anticommuting) fields $\psi_n$ ($n=1,\ldots, {\cal N}$). A constant metric $\eta$ can  be chosen to be
$\eta = diag(1,\ldots,{p_t}^2,\ldots, p_t,\ldots)$ with the $1$ entry repeated $k$ times, the
${p_t}^2$ entry repeated ${\cal N}-k$ times and the $p_t$ entry repeated ${\cal N}$ times.
The global supercharges $Qi$'s, recovered from the classical ones with the prescription introduced before, satisfy  formula (\ref{hermit}). For what concerns the dilatation operator $D$,
the requirement of satisfying (\ref{hermit}) implies constraints on the scaling dimensions
$\lambda_x$, $\lambda_g$ and $\lambda_\psi$ of the component fields $x_l$, $g_m$ and $\psi_n$, respectively. We obtain
\bea
&\lambda_x+{\lambda_x}^\ast=1, \quad \quad\quad
\lambda_g+{\lambda_g}^\ast=-1,\quad\quad\quad
\lambda_\psi+{\lambda_\psi}^\ast=0.&
\eea
The hermiticity condition for the scaling dimension $\lambda_x$ (associated with the metric
$\eta=1$) implies that $\lambda_x$ belongs to the critical strip
\bea
\lambda_x &=& \frac{1}{2}+i\gamma,\quad\quad\quad\quad (\gamma\in {\Bbb R}).
\eea
This is the critical strip where the non-trivial zeros of the Riemann's zeta function are encountered.  This fact is at the core of  a well-known strategy which has been elaborated for proving the Riemann's conjecture by linking it with the hermiticity property of the dilatation operator.  \par
The hermiticity condition implies $\lambda_g$, $\lambda_\psi$ belonging to the strips
$\lambda_g= -\frac{1}{2} +i\gamma'$ and $\lambda_\psi= i\gamma''$ (with $\gamma',\gamma''\in {\Bbb R}$), respectively. By setting the scaling dimensions of the component fields
to be real, it turns out that they differ by $\frac{1}{2}$
($\lambda_\psi=\lambda_g+\frac{1}{2}, \lambda_x=\lambda_\psi+\frac{1}{2}$) as it should be, also in accordance with the classical analysis. \par
The hermiticity conditions depend on the choice of the metric $\eta$, which is not necessarily constant. We illustrate this fact with the example of a single component field $|x>$ with scaling dimension $\lambda$. In the classical framework the real
action (for $\beta$ real)
\bea
{\cal S} &=& \int dt {\cal L}= \int dt(x^\beta{\dot x}^2)
\eea
is scale-invariant and dimensionless provided that the scaling dimension $\lambda$ for the field $x$ satisfies
the condition
\bea\label{scalingdim}
\lambda&=& -{\frac{1}{\beta+2}}
\eea
(the scaling dimension of $t$ is assumed to be $[t]=-1$).\par
Its quantum counterpart is the hermiticity condition
$\int dt <x |\eta|Dx>=\int dt < Dx|\eta |x>$  for a non-constant metric $\eta$ of the form
\bea
\eta &=& A\eta_1+B\eta_2, \quad\quad \eta_1=p_tx^\beta p_t, \quad\quad \eta_2=
{p_t}^2x^\beta+x^\beta {p_t}^2, 
\eea
with $A,B$ some real constants.\par
After straightforward computations, one can show that fulfilling the hermiticity condition implies the vanishing of the coefficients $a,b$ multiplying two types of terms (the only ones surviving after integration by parts), given by
\bea
\left(\int dt <x |\eta|Dx>=\int dt < Dx|\eta |x>\right) &\Longrightarrow & 
\left(a \int dt (x^\beta {\dot x}^2) + b \int dt (tx^{\beta-1}{\dot x}^3 )=0\right).
\eea
The vanishing of $b$ fixes the relative coefficient between $A$ and $B$ to be given
by
\bea\label{AB}
A =-\beta N,\quad &\quad& \quad B= N,
\eea
where $N$ is just a normalization factor.\par
The vanishing of $a$ requires $\lambda$ to satisfy the condition
\bea
\lambda+\lambda^\ast &=& -\frac{2}{\beta+2}.
\eea  
As in the previous cases (for a constant metric $\eta$) we obtain a critical strip. The classical value for the scaling dimension is recovered by requiring $\lambda$ to be real. In the non-constant case the metric $\eta$ has to be conveniently fine-tuned, see formula (\ref{AB}), in order
to obtain non-empty solutions for the hermiticity condition. \par
From this analysis we learn that, for any $\lambda$, the dilatation operator $D$ can be made Hermitian by suitably choosing the metric $\eta$ (specified by the real parameter $\beta$). For superconformal algebras, the hermiticity conditions can be defined in terms of the admissible
metric $\eta$'s or, alternatively, by quantizing the classical real Lagrangians.\par
We are now in position to discuss the criticality conditions (the relation between ${\cal N}=4,7,8$
superconformal algebras and the scaling dimensions,
which coincide with the scaling dimensions of the $x_l$ component fields of their associated global supermultiplets) for Hermitian operators. The $sl(2)$ diagonal operators are expressed
in (\ref{sl2her}), while the global supercharges $Q_i$'s are obtained from the $\frac{d}{dt}\rightarrow p_t$
prescription discussed above. The remaining hermitian generators (the superconformal partners
${\widetilde Q}_i$'s and the $R$-symmetry generators) are determined from the (anti)-commutation relations of the previous generators. For the scaling dimension $\lambda$ defined in (\ref{sl2her}) the criticality conditions coincide with the classical criticality
conditions. The ${\cal N}=8$ superconformal algebras are recovered at $\lambda= \frac{1}{k-4}$
(${\cal N}=7$ at $\lambda=-\frac{1}{4}$)
and the ${\cal N}=4$ relation between $\alpha$ and $\lambda$ is once more given by $\alpha= (2-k)\lambda$.

\section{Superconformal mechanics in Lagrangian framework}

The superconformal algebras that we are dealing with admits the following decomposition in terms
of the grading induced by the dilatation operator $D$ (${\cal G}_i$ is the sector of grading $i$)
\bea
{\cal G} &=& {\cal G}_{-1}\oplus {\cal G}_{-\frac{1}{2}}\oplus
{\cal G}_0\oplus {\cal G}_{\frac{1}{2}}\oplus {\cal G}_{1}.
\eea 
The sector ${\cal G}_1$ (${\cal G}_{-1}$) contains a unique generator given by $H$ ($K$). The odd sectors ${\cal G}_{\frac{1}{2}}$ and ${\cal G}_{-\frac{1}{2}}$ are spanned by the supercharges
$Q_I$'s and their superconformal partners ${\widetilde Q}_I$'s, respectively. The ${\cal G}_0$ sector is given by the union of $D$ and the $R$-symmetry subalgebra (${\cal G}_0={\{D\}}\bigcup\{R\}$).\par
The invariance under the global supercharges $Q_I$'s and the generator $K$ implies the invariance under the full superconformal algebra ${\cal G}$.\par
$D$-module representations can be employed to induce superconformal mechanics in a Lagrangian setting \cite{kt0}. Let the ${\cal N}=4$ supermultiplet $(k, 4, 4-k)$ ($k\geq 1$)
being expressed by $k$ component fields $x_l$ (the propagating bosons), 
$4-k$ auxiliary fields $g_m$ and $4$ fermions $\psi_n$. A global ${\cal N}=4$-invariant action
is obtained from the Lagrangian 
\bea
{\cal L}&=& Q_4Q_3Q_2Q_1[ F(x_l)],
\eea
where $F(x_l)$, known as the {\em prepotential}, is an arbitrary function of the propagating bosons. The ${\cal N}=4$ superconformal invariance  is obtained by suitably constraining $F$, so that the equation
\bea\label{Kcond}
K{\cal L}&=& \frac{d}{dt} M
\eea
($M$ is some function of the component fields and their derivatives) is satisfied. \par
This approach is straightforwardly extended to the multiparticle superconformal mechanics
(based on several ${\cal N}=4$ interacting supermultiplets and such that the prepotential $F$ is a function of all propagating bosons entering the different supermultiplets)  and to ${\cal N}=8$ superconformal mechanics. In this case the ${\cal N}=8$ (with global supercharges $Q_I$, $I=1,\ldots, 8$) $(k,8,8-k)$ supermultiplet  is at first decomposed into
two  ${\cal N}=4$ supermultiplets under $Q_i$'s with $i=1,2,3,4$  ($(k,8,8-k)=(k',4,4-k')\oplus (k-k',4,4-k+k')$). The global ${\cal N}=8$ invariance is obtained by constraining the Lagrangian to satisfy the equations $Q_j{\cal L} =\frac{d}{dt} P_j$, for $j=5,6,7,8$.\par
Further details of this approach to the construction of invariant Lagrangians are found in \cite{kt0}. In \cite{gkt} a slightly more general approach than the one here discussed is presented. It allows to construct ${\cal N}=8$-invariant actions for ${\cal N}=4$ decompositions with $k'=0$ (so that it can be applied to the $(1,8,7)=(1,4,3)\oplus(0,4,4)$ decomposition).  \par
Let us discuss now some applications of the ${\cal N}=8$ critical scaling dimensions we obtained in
this work. We revisit at first the ${\cal N}=8$ $(1,8,7)$ model with a unique propagating boson $x$, fermions $\psi, \psi_j$ and auxiliary fields $g_j$ ($j=1,2,\ldots,7$). Its global ${\cal N}=8$
action has been derived in \cite{krt}. It is given by
\bea\label{187model}
{\cal S}&=& \int dt {\cal L}=\int dt\{(ax+b)[{\dot x}^2-\psi{\dot\psi}-{\dot \psi}_j{\psi_j} +{g_j}^2] +
a[\psi \psi_jg_j-\frac{1}{2}C_{ijk}g_j\psi_j\psi_k]\},
\eea
for some real coefficients $a,b$.
The connection between ${\cal N}=8$ supersymmetry and octonions implies that, without loss of generality, the totally antisymmetric coupling constants $C_{ijk}$ can be identified with the octonionic structure constants. A consistent choice is $C_{123}=C_{147}=C_{165}=C_{257}=C_{354}=C_{367}=1$.\par
It was later shown in \cite{di} that the (\ref{187model}) model can be made superconformally invariant with respect to the $F(4)$ exceptional superalgebra.  The $D$-module analysis of this model
goes as follows. The scale-invariance and the dimensionless of the action requires the homogeneity of the Lagrangian. Therefore, either we have $a=0$ or $b=0$. In the $a=0$ (for $b\neq 0$) case we obtain a constant kinetic term. The scaling dimension $\lambda$ of $x$ coincides with the scaling dimension of the $(1,8,7)$ supermultiplet. It is given, see formula (\ref{scalingdim}) for $\beta=0$, by $\lambda=-\frac{1}{2}$. This value, however, does not coincide with the critical scaling dimension for the ${\cal N}=8$ supermultiplet with $k=1$. In the second case ($b=0$ and $a\neq0$) we obtain a nontrivial Lagrangian, due to the presence of the cubic term. The scaling dimension $\lambda$ is now recovered from (\ref{scalingdim}) with $\beta=1$. We obtain for this value the critical scaling dimension $\lambda=-\frac{1}{3}$ 
of the ${\cal N}=8$ $k=1$ supermultiplet. At this critical value the $(1,8,7)$ supermultiplet induces a $D$-module representation of the $F(4)$ superconformal algebra. Straightforward computations show that, at $\lambda=-\frac{1}{3}$, the action of the $K$ generator on the (\ref{187model}) Lagrangian satisfies the (\ref{Kcond}) condition. We recover, with a different method, the $F(4)$ superconformal invariance of the (\ref{187model}) model for $b=0$. Unlike the superspace approach of \cite{di}, the $F(4)$ generators act linearly on the $(1,8,7)$ component fields. \par
The next model we analyze is based on the inhomogeneous ${\cal N}=8$ $(2,8,6)$ supermultiplet
and was introduced in \cite{lt}. In the Lagrangian $D$-module approach the inhomogeneity
(expressed by a real parameter $c$)
is essential to produce Calogero-type terms in the action.  The presence of the inhomogeneous term in the global ${\cal N}=8$ transformations implies that the $6$ auxiliary fields have the same scaling
dimension ($=0$) of the real inhomogeneous parameter $c$. This requirement unambiguosly fixes the scaling dimension of the $(2,8,6)$ inhomogeneous supermultiplet to be $\lambda=-1$.
The present analysis proves that $\lambda=-1$ does not coincide with the critical scaling
dimension of the ${\cal N}=8$ $k=2$ case. As a consequence the scale-invariant, global ${\cal N}=8,$ model of \cite{lt} does not possess a superconformal invariance under a finite
superconformal algebra.\par
Quite a different picture is recovered for the uniquely defined scale-invariant and global ${\cal N}=8$ model based on the inhomogeneous $(3,8,5)$ supermultiplet (the only arbitrariness is the value of the inhomogeneous parameter $c$). For $k=3$, $\lambda=-1$ is a critical scaling dimension. It can be proven that the action of this model, derived from the inhomogeneous $(3,8,5)$ ${\cal N}=8$ transformations introduced in subsection {\bf 3.6},  is invariant under the $D(2,2)=osp(4|4)$ superconformal algebra.  Its explicit presentation and the discussion of its properties is left to a forthcoming paper in preparation.

 \section{The $S_3$ $\alpha$-orbit of $D(2,1;\alpha)$ and the constraints on multiparticle
superconformal mechanics}

The finite ${\cal N}=4$ simple superconformal algebras are $A(1,1)$ and the exceptional superalgebras $D(2,1;\alpha)$, for $\alpha\in {\Bbb C}\backslash\{0,-1\}$ \cite{dictionary}. The superalgebras
$D(2,1;\alpha)$'s are isomorphic if and only if the parameters $\alpha$ are connected via an
$S_3$ group of transformations generated by the moves $\alpha\mapsto \frac{1}{\alpha}$ and
$\alpha \mapsto - (1+\alpha)$. We have therefore at most $6$ different $\alpha$'s producing,
up to isomorphism,  the superconformal algebra $D(2,1;\alpha)$. They are given, explicitly, by 
\bea\label{alphas}
&
\begin{array}{lllllllll}
\alpha^{(1)} &=&\alpha,\quad&\alpha^{(3)}&=&-(1+\alpha),\quad& \alpha^{(5)}&=&-\frac{1+\alpha}{\alpha},\\
\alpha^{(2)} &=&\frac{1}{\alpha},\quad&\alpha^{(4)}&=&-\frac{1}{(1+\alpha)},\quad& \alpha^{(6)}&=&-\frac{\alpha}{(1+\alpha)}.
\end{array}
&
\eea
It is convenient to regard $A(1,1)$ as a degenerate superalgebra recovered from $D(2,1;\alpha)$
at the special values $\alpha=0,-1$ (at these special values three even generators decouple from the rest of the generators; the remaining ones close the $A(1,1)$ superalgebra).\par
The inequivalent ${\cal N}=4$ simple superconformal algebras are therefore expressed by the fundamental domain obtained by quotienting the complex plane ($\alpha \in {\Bbb C}$) under the action of the $S_3$ group. In application to classical superconformal mechanics, $\alpha$ is restricted to be real ($\alpha\in\Bbb{R}$). With this restriction a fundamental domain under the action of the $S_3$ group can be chosen to be the closed interval  
\bea{\label{funddom}}
\relax \alpha &\in& [0,1].
\eea 
Some points in the interval are of special significance. We have that\\
{\em i)} - the extremal point $\alpha=0$ corresponds to the $A(1,1)$ superalgebra,\\
{\em ii)} - the extremal point $\alpha =1$ correspond to the $D(2,1)$ superalgebra, belonging
to the $D(m|n)=osp(2m|2n)$ classical series,\\
{\em iii)} - the midpoint $\alpha=\frac{1}{2}$ corresponds to the $F(4)$ subalgebra
$D(2,1;\frac{1}{2})\subset F(4)$,\\
{\em iv)} - the point $\alpha=\frac{1}{3}$ corresponds to the $G(3)$ subalgebra
$D(2,1;\frac{1}{3})\subset G(3)$.\par
We will see in the following the special role played by these points.
\par
The combined properties of having different $\alpha$'s producing isomorphic ${\cal N}=4$ superconformal algebras (\ref{alphas}), together with the set of critical relations between $\alpha$ and the scaling dimension $\lambda$  of the $(k, 4, 4-k)$ ${\cal N}=4$ supermultiplets (with $k=0,1,2,3,4$), given by
\bea
\alpha &=& (2-k)\lambda,
\eea
produce highly non-trivial constraints on the admissible ${\cal N}=4$ superconformal mechanics models and their scaling dimension. For the $\alpha=0,-1$ case, e.g., we have that the solutions are recovered for any real $\lambda$ for the $(2,4,2)$ supermultiplet (for $\lambda\neq 0$ the superalgebra is $sl(2|2)$, see \cite{kt0}) while, for $k\neq 2$, they are obtained for $\lambda=0$ or $\lambda=\frac{1}{k-2}$. It is convenient to summarize some results in a table presenting the admissible 
scaling dimension $\lambda$ associated to the $(k,4,4-k)$ supermultiplets for the above four
cases, specified by $\alpha_{FD}$ (the value $\alpha$ in the (\ref{funddom}) fundamental domain) given by,
respectively, $\alpha_{FD}=0,1,\frac{1}{2},\frac{1}{3}$. We have
\par
\quad\par
\quad\quad\quad\quad\quad\quad
\begin{tabular}{|cc|}\hline
$\alpha_{FD}=0$:&
\begin{tabular}{|l|c|}
 $k$&$\lambda $ \\
\hline
 $0$& $0,-\frac{1}{2}$\\ \hline
  $1$& $0,-1$\\ \hline
  $2$& $\Bbb{R}$\\ \hline
  $3$& $0,1$\\ \hline
  $4$& $0,\frac{1}{2}$\\ 
\end{tabular}\\ \hline
\end{tabular}\par
\quad\quad\quad\quad\quad\quad
\begin{tabular}{|cc|}\hline
$\alpha_{FD}=1$:&
\begin{tabular}{|l|c|}
 $k$&$\lambda $ \\
\hline
 $0$& $-1,-\frac{1}{4}, \frac{1}{2}$\\ \hline
  $1$& $-2,-\frac{1}{2},1$\\ \hline
  $3$& $-1,\frac{1}{2},2$\\ \hline
  $4$& $-\frac{1}{2},\frac{1}{4},1$\\ 
\end{tabular}\\ \hline
\end{tabular}\par
\quad\quad\quad\quad\quad\quad
\begin{tabular}{|cc|}\hline
$\alpha_{FD}=\frac{1}{2}$:&
\begin{tabular}{|l|c|}
 $k$&$\lambda $ \\
\hline
 $0$& $ -\frac{3}{2},-\frac{3}{4},-\frac{1}{3},-\frac{1}{6},\frac{1}{4},1 $\\ \hline
  $1$& $-3, -\frac{3}{2},-\frac{2}{3},-\frac{1}{3},\frac{1}{2},2$\\ \hline
  $3$& $-2,-\frac{1}{2}, \frac{1}{3}, \frac{2}{3},\frac{3}{2}, 3$\\ \hline
  $4$& $-1, -\frac{1}{4},\frac{1}{6},\frac{1}{3},\frac{3}{4},\frac{3}{2}$\\ 
\end{tabular}\\ \hline
\end{tabular}\par
\quad\quad\quad\quad\quad\quad
\begin{tabular}{|cc|}\hline
$\alpha_{FD}=\frac{1}{3}$:&
\begin{tabular}{|l|c|}
 $k$&$\lambda $ \\
\hline
 $0$& $  -2, -\frac{2}{3}, -\frac{3}{8},-\frac{1}{8}, \frac{1}{6},\frac{3}{2} $\\ \hline
  $1$& $ -4, -\frac{4}{3}, -\frac{3}{4}, -\frac{1}{4},\frac{1}{3}, 3$\\ \hline
  $3$& $
-3,-\frac{1}{3}, \frac{1}{4}, \frac{3}{4}, \frac{4}{3}, 4$\\ \hline
  $4$& $
-\frac{3}{2},-\frac{1}{6},\frac{1}{8}, \frac{3}{8}, \frac{2}{3},2$\\ 
\end{tabular}\\ \hline
\end{tabular}
\begin{eqnarray}\label{rcase}&
&\end{eqnarray}
The ${\cal N}=4$ superconformal invariance for several (at least two,
let's say $(k,4,4-k)$ and $(k',4,4-k')$) interacting supermultiplets requires that they should carry a $D$-module representation for the
same $D(2,1;\alpha)$ superalgebra. 
Given two supermultiplets with $k'\neq k$, this requirement  produces
{\em a)}  a constraint on the admissible values for $\alpha$ and
{\em b)} a consequent constraint on the mutual scaling dimensions of the two supermultiplets
(both these constraints will be called the ``compatibility condition").\par
The table above shows that, for $\alpha_{FD}=0,1,\frac{1}{2}$, two supermultiplets with $k'\neq k$ and the same scaling dimension $\lambda$ can be found for
\begin{eqnarray}&
\begin{tabular}{lll}\label{compatalpha}
$\alpha_{FD}=0:$& $\lambda =0 \quad \quad \quad\quad (k,k'\neq 2),\quad$&$\lambda = 0,\frac{1}{k-2} \quad ( k'=2),$\\
$\alpha_{FD}=1:$& $\lambda =-1,\frac{1}{2} \quad \quad (k=0, k'=3),\quad$&$ \lambda = 1,-\frac{1}{2} \quad ( k=1, k'=4),$\\
$\alpha_{FD}=\frac{1}{2}:$& $\lambda =-\frac{3}{2},-\frac{1}{3} \quad (k=0, k'=1),\quad$&$ \lambda = \frac{3}{2},\frac{1}{3} \quad \quad( k=3, k'=4).$
\end{tabular}
&\end{eqnarray}
Some conclusions can be drawn from this result. For instance, the decomposition of ${\cal N}=8$ $D$-module representations into ${\cal N}=4$ supermultiplets can (partly) explain the ${\cal N}=8$ critical scaling dimensions. The $(7,8,1)$ supermultiplet gets decomposed into $(4,4,0)\oplus (3,4,1)$. Its critical scaling dimension is therefore constrained to be either $\lambda=\frac{3}{2}$ or $\lambda=\frac{1}{3}$ (its actual value). The $(4,4,0)\oplus (2,4,2)$ decomposition of the $(6,8,2)$ supermultiplet implies that its critical scaling dimension can
only be found at $\lambda=0$ or $\lambda=\frac{1}{2}$  (its actual value). The $(5,8,3)$ supermultiplet admits the decompositions $(4,4,0)\oplus (1,4,3)$ and $(3,4,1)\oplus (2,4,2)$.
Their combination uniquely implies a possible critical scaling dimension at $\lambda=1$ (its actual value).\par
The $\alpha_{FD}=\frac{1}{3}$ case does not admit a common scaling dimension $\lambda$ if $k\neq k'$. This value of $\alpha$ corresponds to the decomposition of the ${\cal N}=7$ $(1,7,7,1)$ supermultiplet into the ${\cal N}=4$ $(1,4,3,0)\oplus(0,3,4,1)$ supermultiplets. 
Let us denote with $\lambda_1,\lambda_3$ the respective scaling dimensions of these ${\cal N}=4$ supermultiplets. Clearly it must be $\lambda_3=\lambda_1+\frac{1}{2}$. An inspection of the
(\ref{rcase}) table shows that the unique pair of values differing by $\frac{1}{2}$ are recovered for
$\lambda_1=-\frac{1}{4}$ and $\lambda_3=\frac{1}{4}$. This analysis corroborates the finding of
$\lambda=-\frac{1}{4}$ as the critical scaling dimension of the $(1,7,7,1)$ $D$-module representation of $G(3)$.\par
We are also able to partly explain the arising of ${\cal N}=6$ superconformal algebras from length-$4$ supermultiplets, obtained in Section {\bf 4}. The $(2,6,6,2)$ supermultiplet
produces the ${\cal N}=6$ superconformal algebra $A(2,1)$ for any value of $\lambda$ (no criticality). Its ${\cal N}=4$ decomposition reads as $(2,6,6,2)=(2,4,2,0)\oplus (0,2,4,2)$. One should note, from table (\ref{rcase}), that no restriction on $\lambda$ is put from the $(2,4,2)$ supermultiplets. On the other hand the $(1,6,7,2)$ supermultiplet induces the $A(2,1)$ superalgebra at the critical value $\lambda=0$. The ${\cal N}=4$ decomposition is expressed
as $(1,6,7,2)=(1,4,3,0)\oplus(0,2,4,2)$. The presence of both supermultiplets $(1,4,3)$ and $(2,4,2)$ requires $\alpha=0,-1$. The admissible values of critical $\lambda$, obtained from this analysis, are $\lambda=0,-1$.  The $(2,7,6,1)$ supermultiplet induces the $A(2,1)$ superalgebra at the critical value $\lambda=-\frac{1}{2}$. Let $\lambda_2$, $\lambda_3$ be the scaling dimensions of the respective ${\cal N}=4$ supermultiplets entering the ${\cal N}=4$ decomposition $(2,7,6,1)=(2,4,2,0)\oplus(0,3,4,1)$.  $\lambda_2$ coincides with the scaling dimension $\lambda$ of the $(2,7,6,1)$ supermultiplet, while $\lambda_3=\lambda_2+\frac{1}{2}$. The $\alpha=0,-1$ constraint on $\lambda_3$ ($\lambda_3=0,1$)  implies that in this case the critical $\lambda$ is constrained (necessary condition) to be $\lambda=\pm\frac{1}{2}$. 
\par
Let us deal now with the general case of finding the compatibility conditions on $\alpha$ and the common scaling dimension $\lambda$ for two ${\cal N}=4$ $D$-module representations with $k\neq k'$. It is sufficient to discuss the $k,k'\neq 2$ restriction, since the remaining cases are immediately recovered from the $\lambda$ solutions at $\alpha=0,-1$.  Without loss of generality we can set $\alpha\equiv \alpha^{(1)}=(2-k)\lambda$ for the $k$ supermultiplet. The $\alpha'$ value
obtained as $\alpha'=(2-k')\lambda$ from the $k'$ supermultiplet must coincide with one of the $\alpha^{(i)}$ in the $S_3$-orbit of $\alpha$. Let us introduce the ratios
\bea
N^{(i)} &=& \frac{\alpha^{(i)}}{\alpha^{(1)}}
\eea
and
\bea
w_{kk'}&=& \frac{2-k'}{2-k}= \frac{1}{w_{k'k}}.
\eea
The values obtained by $w_{kk'}$ in varying $k,k'$ with the given constraints are $-1, \pm \frac{1}{2}, \pm 2$.\par
For a given pair $[k,k']$ the admissible values $\alpha$ satisfying the compatibility condition 
are recovered by ${\overline \alpha}$ and its $S_3$-group orbit
(\ref{alphas}), with ${\overline\alpha}$ a solution of one of the five equations (for $i=2,3,4,5,6$,
since $N^{(1)}=w$ has no solution for $w\neq 1$) 
\bea\label{nwsystem}
N^{(i)}&=& w
\eea
(since no confusion will arise, for simplicity, we set $w\equiv w_{kk'}$).\par
The compatibility conditions are recovered from the (\ref{nwsystem}) system of equations for three inequivalent values of $w$, given by $w=-1,-2,2$.  This is due to the fact that the transformation
$w\leftrightarrow\frac{1}{w}$ reflects the $k\leftrightarrow k'$ exchange.\par 
Two of the (\ref{nwsystem}) equations are linear in $\alpha$, while the three remaining ones are quadratic (producing, in some cases, complex solutions).\par
The ${\overline\alpha}$ solutions can be divided into three classes: real and rational, real and irrational, complex. \par
The complex solutions (associated to scaling dimension $\lambda$'s which do not satisfy the reality condition) are found to be
\bea
w=-1 &:& {\overline \alpha}=\pm i,\nonumber\\
w=-2 &:& {\overline \alpha}=\pm \frac{i}{\sqrt{2}},\nonumber\\
w= 2 &:& {\overline \alpha}=\frac{1}{2}(-1\pm i), -1\pm i{\sqrt{7}}.
\eea
 For what concerns the real solutions the following results are obtained.\par
In the rational case, the unique solutions are encountered for $w=-2$ (the ${\overline\alpha}$ $S_3$-orbit is specified by $\alpha_{FD}=1$) and $w=2$ (with orbit specified by $\alpha_{FD}=\frac{1}{2}$). We therefore recover the solutions already encountered in
(\ref{compatalpha}) and their corresponding scaling dimension 
 $\lambda$'s. No further rational solution is found.\par
For what concerns the irrational case the encountered results are summarized in the table below, which specifies the $[k,k']$ pairs, the value $w$, the $S_3$ orbit representative $\alpha_{FD}$ in the fundamental domain and, finally, the compatible scaling dimension $\lambda$'s. We have
\begin{center}
\begin{tabular}{|c|c|c|c|}\hline
\relax $[k,k']$& $w$&$\alpha_{FD}$& $\lambda $ \\
\hline\hline
$[1,3]$&  $-1$& $-\frac{1}{2}+\frac{\sqrt{5}}{2} $ & $-\frac{1}{2}\pm\frac{\sqrt{5}}{2} $, $\frac{1}{2}\pm\frac{\sqrt{5}}{2}$\\ \hline
$[0,4]$&  $-1$& $-\frac{1}{2}+\frac{\sqrt{5}}{2} $ & $-1\pm{\sqrt{5}}$, $1\pm\sqrt{5}$\\ \hline \hline
$[1,4]$&  $-2$& $-\frac{1}{2}+\frac{\sqrt{3}}{2}$ & $-\frac{1}{2}\pm\frac{\sqrt{3}}{2} $\\ \hline
$[3,0]$&  $-2$& $-\frac{1}{2}+\frac{\sqrt{3}}{2}$ & $\frac{1}{2}\pm\frac{\sqrt{3}}{2}$\\ \hline \hline
$[1,0]$&  $2$& $\frac{1}{\sqrt{2}}$ & $\sqrt{2}$\\ \hline
$[3,4]$&  $2$& $\frac{1}{\sqrt{2}}$ & $-\sqrt{2}$\\ \hline
$[1,0]$&  $2$& $\sqrt{2}-1$&$-\sqrt{2}$\\ \hline
$[3,4]$&  $2$& $\sqrt{2}-1$ & $\sqrt{2}$\\ \hline
\end{tabular}
\end{center}
\begin{eqnarray}\label{irrcase}&&\end{eqnarray}
These results give the necessary condition for the existence of ${\cal N}=4$ superconformal mechanics based on several inequivalent interacting supermultiplets with the same scaling dimension $\lambda$.\par
Unlike the $\alpha_{FD}=1$ with $\lambda=1$  (for $[k,k']=[1,4]$) and $\lambda=-1$ (for $[k,k']=[0,3]$) and $\alpha_{FD}=\frac{1}{2}$ with $\lambda=\frac{1}{3}$ (for $[k,k']=[3,4]$) and
$\lambda = -\frac{1}{3}$ (for $[k,k']=[0,1]$), the irrational cases and the remaining rational cases 
do not allow the extension of the ${\cal N}=4$ superconformal invariance to a broader ${\cal N}=8$ superconformal invariance.\par
A particularly interesting case involves the irrational solution of the $k=1$, $k'=3$ supermultiplets. The value $\alpha_{FD}$ coincides with the golden mean conjugate $\Phi=-\frac{1}{2}(1-\sqrt{5})$
(the golden mean $\varphi=\frac{1}{2}(1+\sqrt{5})$ belongs to its $S_3$-orbit). This case admits four compatible solutions for the scaling dimension $\lambda$ (given by $\pm \varphi$ and $\pm \Phi$).
${\cal N}=4$ superconformal actions, invariant under $D(2,1;\alpha=\varphi)$, are obtained
for the pairs of ${\cal N}=4$ supermultiplets $(x;\psi,\psi_i; g_i)$ and $(y_i;\xi,\xi_i;h)$,
$i=1,2,3$, in terms of the
Lagrangians ${\cal L} = Q_4Q_3Q_2Q_1 F(x, y_i)$. The $x,y_i$ fields are the propagating bosons. A class of solutions, satisfying the (\ref{Kcond}) constraint, is obtained for the prepotential $F(x,y_i)$ given by 
\bea\label{goldenprep}
F(x,y_i) &=& Cx^\beta r^\gamma,\quad \quad \beta+\gamma=\frac{1}{\varphi}
\eea
(here $r=\sqrt{y_1^2+y_2^2+y_3^2}$ and $C$ is an arbitrary constant) and linear combinations thereof.

\section{Conclusions}

In this work we constructed a class of $D$-module representations for one-dimensional superconformal algebras. These representations exhibit, for ${\cal N}=4,7,8$, the property of criticality. This means that they only close at critical values
of the scaling dimension $\lambda$ characterizing the supermultiplets of time-dependent component fields. The superalgebras under consideration are a given subclass of finite, simple,  Lie superalgebras. Their $D$-module representations are an extension of the $D$-module representation (\ref{sl2dmod})  of the $sl(2)$ algebra (this representation is non-critical, being recovered for any value of $\lambda$). \par
The connection with the $D$-module representations \cite{krt} of the ${\cal N}$-Extended global supersymmetry (the superalgebra of the one-dimensional Supersymmetric Quantum Mechanics)
is given by the fact that the latter is a subalgebra of the superconformal algebras.
Certain minimal global supermultiplets induce, at a given $\lambda$, their associated $D$-module representations of a superconformal algebra. In particular, the exceptional finite Lie superalgebras
\cite{{kac},{dictionary}}
$D(2,1;\alpha)$, $G(3)$ and $F(4)$ (which are superconformal algebras for ${\cal N}=4,7,8$, respectively) admit a $D$-module representation. Quite interestingly, the $D$-module representation of $G(3)$ is only induced from the minimal $(1,7,7,1)$ global ${\cal N}=7$ supermultiplet, first introduced in \cite{krt}.\par
The action of the generators of a $D$-module representation on a set of component fields allows to construct superconformal mechanical models in a Lagrangian framework. 
With our analysis based on $D$-module representations we have been able to prove the invariance of certain 
superconformal models, to check that the \cite{lt} model is not invariant under a finite simple Lie superalgebra while, on the other hand, we pointed out the existence of a $D(2,2)$-invariant superconformal model based on the inhomogeneous $(3,8,5)$ supermultiplet.
\par
An updated review on several issues of superconformal mechanics (including application to test particles moving near black hole horizons, $CFT_1/AdS_2$ correspondence, etc.) can be found in \cite{fil} (see also the references therein).\par
The isomorphism of the $D(2,1;\alpha)$ superalgebras under an $S_3$ group of transformations acting on the parameter $\alpha$, together with the relation between $\alpha$ and $\lambda$
for the $(k,4,4-k)$ ${\cal N}=4$ supermultiplets, gives non-trivial restrictions on the
admissible ${\cal N}=4$ multi-particle superconformal mechanics (based on several interacting
supermultiplets). In Appendix {\bf B} we presented the constraints on the scaling dimensions
of the superconformal models with interacting supermultiplets.
We proved in particular the existence of an ${\cal N}=4$ superconformal mechanics (see formula (\ref{goldenprep})) for the interacting $(1,4,3)$ and $(3,4,1)$ supermultiplets, with scaling dimension based on the golden ratio. \par
The representation theoretical nature of the constraints on multi-particle ${\cal N}=4$ superconformal mechanics allows to partly explain the critical scaling dimensions $\lambda$ 
encountered at ${\cal N}=7$ and ${\cal N}=8$. This is due to the fact that the
${\cal N}=7,8$ supermultiplets admit (at least one) decomposition into two separate ${\cal N}=4$ supermultiplets.\par
A natural extension of this work consists in investigating $D$-module representations for
twisted superconformal algebras. A $D$-module representation for a twisted version of the ${\cal N}=2$ superconformal algebra was constructed in \cite{bt}. The investigation of $D$-modules for larger values of (twisted) ${\cal N}$ superconformal algebras seems a promising tool to analyze the dimensional reduction (to one dimension) of the ${N}=4$ super-Yang-Mills theory.

~\\{~}~
{\Large{\bf Appendix A: on global supermultiplets}}\\
~\par
For completeness we recall here the features of the global supermultiplets of the one-dimensional ${\cal N}$-Extended superalgebra that are relevant for this work. The general theory
is discussed in \cite{krt}. The linear, homogeneous, minimal supermultiplets contain $4$ bosonic and $4$ fermionic component fields for ${\cal N}=3,4$ and $8$ bosonic and $8$ fermionic component fields for ${\cal N}=5,6,7,8$ (the non-minimal ${\cal N}=4$ supermultiplets also contain $8$ bosonic and $8$ fermionic fields).\par
The minimal ${\cal N}=3,4,7,8$ supermultiplets are uniquely characterized by their
``field content" $(n_1,n_2,\ldots, n_l)$, expressing the number $n_i$ of component fields with  scaling dimension $\lambda_i = \lambda+ \frac{i-1}{2}$. The index $l$, known as the length of the supermultiplet, denotes the number of inequivalent scaling dimensions in the supermultiplet.
The bosonic fields are associated to odd values of $i$, the fermionic fields to even values. The equality
$n_1+n_3+\ldots = n_2+n_4+\ldots=n$ holds, with $n=4$ for ${\cal N}=3,4$ and $n=8$ for ${\cal N}=7,8$.\par
In particular the $(k,n,n-k)$ length-$3$ supermultiplet admits a set of $k$ bosonic fields
$x_j(t)$ ($j=1,2,\ldots, k$), $n$ fermionic fields $\psi_a(t)$ ($a=1,2,\ldots,n$) and $(n-k)$
bosonic fields $g_m(t)$ ($m=1,2,\ldots, n-k$). All fields depend on the real parameter $t$ (the ``time"). Their respective scaling dimensions are $[x_j(t)]=\lambda$, $[\psi_a(t)]=\lambda+\frac{1}{2}$, $[g_m(t)]=\lambda+1$ ($\lambda$ is the overall scaling dimension of the supermultiplet). For their role in the one-dimensional supersymmetric sigma-models, the $x_j$ fields will be
called the {\em propagating bosons} or {\em target coordinates}, while the $g_m$ fields will be called {\em auxiliary fields}.  \par
The length-$2$ supermultiplets $(n,n)$, known as ``root supermultiplets", are of particular relevance. They are uniquely determined \cite{pt} by their associated Clifford algebra representation. 
Their supertransformations can be encoded in $2{ n}\times 2{n}$ supermatrices whose entries are differential operators (the non-vanishing entries are either $\pm1$ or $\pm \frac{d}{dt}$).
The remaining
supermultiplets of higher length are obtained from the given root supermultiplet via a dressing transformation determined by a dressing operator $S$. $S$ is a diagonal differential operator whose diagonal entries (for the cases here considered) are $1$ and $\partial_t=\frac{d}{dt}$.  \par
Let $Q_I^R$ be the supermatrices expressing the supercharges in the root representation. The supermatrices ${Q_I}$ of a dressed representation are given by
\bea\label{qdressing}
Q_I&=& SQ_I^RS^{-1}.
\eea
One should note that $S^{-1}$ is a pseudo-differential operator. The requirement that the supermatrices $Q_I$ are differential operators (see the analysis in \cite{{pt},{krt}}) puts a constraint on the admissible dressings and, as a consequence, on the admissible higher length
supermultiplets. \par
Without loss of generality we used the following conventions for the ${\cal N}=8$ root supermultiplet.  The $8$ supercharges are given by 
\bea\label{n8scharges}
 Q_J^R&=& \left(
\begin{array}{cc}
0&\gamma_J\\
-\gamma_J \partial_t&0
\end{array}
\right),\nonumber\\
 Q_8^R& =&\left(
\begin{array}{cc}
0&{\bf 1}_8\\
{\bf 1}_8 \partial_t&0
\end{array}
\right),
\eea
where  the $\gamma_J$ matrices ($J=1,\ldots , 7$) are the generators of the $Cl(0,7)$ Euclidean Clifford algebra: $\{\gamma_J,\gamma_L\} = -2\delta_{JL}{\bf 1}_8$. We have, explicitly,
\bea
&&\gamma_{1} = \tau_{2}\otimes\tau_{2}\otimes \tau_{3},\quad 
\gamma_{2} = \tau_{2}\otimes\tau_{3}\otimes \textbf{1}_{2},\quad
\gamma_{3} = \tau_{2}\otimes\tau_{1}\otimes \tau_{3}, \nonumber\\
&&\gamma_{4} = \tau_{3}\otimes\textbf{1}_{2}\otimes \textbf{1}_{2},\quad
\gamma_{5} = \tau_{1}\otimes\textbf{1}_{2}\otimes\tau_{3},\quad
\gamma_{6} = \tau_{1}\otimes\tau_{3}\otimes \tau_{2},\nonumber\\
&&\gamma_{7} = \tau_{1}\otimes\tau_{3}\otimes \tau_{1}.
\eea
The three $2\times 2$ matrices $\tau_{i}$ are 
$\tau_1 = \left(
\begin{array}{cc}
0&1\\
1&0
\end{array}
\right)$, $\tau_2 = \left(
\begin{array}{cc}
1&0\\
0&-1
\end{array}
\right)$ $\tau_3 = \left(
\begin{array}{cc}
0&1\\
-1&0
\end{array}
\right)$.\par
The ${\cal N}=8$ minimal global supermultiplets $(k,8,8-k)$ admits an inhomogeneous extension for $k=0,1,2,3$. Among them, the $(3,8,5)_{inhom}$ global inhomogeneous
supermultiplet is the only one which induces a $D$-module representation for an ${\cal N}=8$
superconformal algebra.

~\\{~}~\\
{\Large{\bf Appendix B: admissible common real scaling dimension for pairs of ${\cal N}=4$ superconformal multiplets}}\\
~\par
We report here for clarity a table containing the admissible, common, {\em real} values of the scaling dimension $\lambda$ for pairs of ${\cal N}=4$ superconformal multiplets transforming under the same ${\cal N}=4$ superconformal algebra. This information, extracted from the tables (\ref{rcase}) and (\ref{irrcase}), is important in constraining the 
admissible multiparticle superconformal mechanics. We have
\bea
(4,4,0)\oplus (4,4,0) &:& \lambda \in {\Bbb R},\nonumber\\
(4,4,0)\oplus (3,4,1) &:& \lambda=\frac{3}{2}, \frac{1}{3}, \pm \sqrt{2},\nonumber\\
(4,4,0)\oplus (2,4,2) &:& \lambda =0,\frac{1}{2},\nonumber\\
(3,4,1)\oplus (3,4,1) &:& \lambda \in {\Bbb R},\nonumber\\
(4,4,0)\oplus (1,4,3)&:& \lambda =1, -\frac{1}{2}, -\frac{1}{2}\pm \frac{\sqrt{3}}{2},\nonumber\\
(3,4,1)\oplus (2,4,3) &:& \lambda =0,1,\nonumber\\
(4,4,0)\oplus (0,4,4) &:& \lambda=-1\pm\sqrt{5}, 1\pm\sqrt{5},\nonumber\\
(3,4,1)\oplus (1,4,3) &:& \lambda =-\frac{1}{2}\pm\frac{{\sqrt{5}}}{2},\frac{1}{2}\pm\frac{\sqrt{5}}{2}\nonumber\\
(2,4,2)\oplus (2,4,2) &:& \lambda \in {\Bbb R},\nonumber\\
(3,4,1)\oplus (0,4,4)&:& \lambda =-1, \frac{1}{2}, \frac{1}{2}\pm \frac{\sqrt{3}}{2},\nonumber\\
(2,4,2)\oplus (1,4,3) &:& \lambda =0,-1,\nonumber\\
(2,4,2)\oplus (0,4,4) &:& \lambda=0, -\frac{1}{2},\nonumber\\
(1,4,3)\oplus (1,4,3) &:& \lambda \in {\Bbb R},\nonumber\\
(1,4,3)\oplus (0,4,4) &:& \lambda = -\frac{3}{2}, -\frac{1}{3}, \pm \sqrt{2},\nonumber\\
(0,4,4)\oplus (0,4,4)&:& \lambda \in {\Bbb R}.\nonumber\\
\eea
This table allows to explain why there is no ${\cal N}=8$ superconformal $D$-module based
on the $(4,8,4)$ supermultiplet. Indeed, depending on the choice of the ${\cal N}=4$ subalgebra, this supermultiplet admits the three ${\cal N}=4$ decompositions 
$(4,4,0)\oplus (0,4,4)$, $(3,4,1)\oplus (1,4,3)$ and $(2,4,2)\oplus (2,4,2)$. There is, however, no common scaling dimension $\lambda$ belonging to both $(4,4,0)\oplus (0,4,4)$ and $(3,4,1)\oplus (1,4,3)$.\par
One should note that the $(k,n,n-k)\leftrightarrow (n-k,n,k)$ ``mirror duality" of the global supermultiplets (see \cite{krt}) extends to a duality for the scaling dimension of the superconformal multiplets.
~{}
\\ {~}~
\par {\large{\bf Acknowledgments}}
{}~\par~\par
We are grateful to M. Gonzales and Z. Kuznetsova for useful discussions. This work was supported by CNPq.

\end{document}